# Frontier AI developers need an internal audit function

*Jonas Schuett\**

This article argues that frontier artificial intelligence (AI) developers need an internal audit function. First, it describes the role of internal audit in corporate governance: internal audit evaluates the adequacy and effectiveness of a company's risk management, control, and governance processes. It is organizationally independent from senior management and reports directly to the board of directors, typically its audit committee. In the IIA's Three Lines Model, internal audit serves as the third line and is responsible for providing assurance to the board, while the Combined Assurance Framework highlights the need to coordinate the activities of internal and external assurance providers. Next, the article provides an overview of key governance challenges in frontier AI development: dangerous capabilities can arise unpredictably and undetected; it is difficult to prevent a deployed model from causing harm; frontier models can proliferate rapidly; it is inherently difficult to assess frontier AI risks; and frontier AI developers do not seem to follow best practices in risk governance. Finally, the article discusses how an internal audit function could address some of these challenges: internal audit could identify ineffective risk management practices; it could ensure that the board of directors has a more accurate understanding of the current level of risk and the adequacy of the developer's risk management practices; and it could serve as a contact point for whistleblowers. But frontier AI developers should also be aware of key limitations: internal audit adds friction; it can be captured by senior management; and the benefits depend on the ability of individuals to identify ineffective practices. In light of rapid progress in AI research and development, frontier AI developers need to strengthen their risk governance. Instead of reinventing the wheel, they should follow existing best practices. While this might not be sufficient, they should not skip this obvious first step.

## 1. Introduction

The past few years have shown a remarkable trend: using more data and more compute to train bigger artificial intelligence (AI) models leads to predictable improvements in their performance.[1] This phenomenon is commonly referred to as "scaling laws" (Hestness et al., 2017; Kaplan et al., 2020; Bahri et al., 2021; Hoffmann et al., 2022), and the claim that this trend will continue as the "scaling hypothesis" (Gwern, 2020). While it is unclear if the scaling hypothesis is true (Lohn & Musser, 2022; Villalobos et al., 2024a; 2024b; Narayanan & Kapoor, 2024), the underlying trend has already led to the development of highly capable and increasingly general AI systems, such as GPT-4o (OpenAI, 2024c), Claude 3.5 Sonnet (Anthropic, 2024b, 2024c), and Gemini 1.5 (Google DeepMind, 2024b).

---

\* Centre for the Governance of AI, Oxford, UK. Email: jonas.schuett@governance.ai.

[1] More precisely, we can empirically observe that increasing the size of the training dataset ($D$), the number of parameters ($P$), and the amount of training compute ($C$) leads to a predictable reduction in the test loss ($L$), i.e. the difference between the model's predicted output and the actual target value on an unseen dataset. The relationship between the properties $D$, $P$, $C$, and $L$ follows a precise power law (Kaplan et al., 2020; Bahri et al., 2021).

But this trend also has concerning implications. As models are scaled up, new capabilities can emerge unintentionally and unpredictably (Ganguli et al., 2022a; Wei et al., 2022),[2] some of which might be dangerous (Shevlane et al., 2023; Phuong et al., 2024). For example, models might become able to deceive, persuade, and manipulate people (Park et al., 2023; Carlsmith, 2023; Hubinger et al., 2024; El-Sayed et al., 2024; Hackenburg & Margetts, 2024; Hagendorff, 2024), find and exploit cyber vulnerabilities (Mirsky et al., 2021; Lohn & Jackson, 2022; Fang et al., 2024; Zhang et al., 2024), or provide instructions for the acquisition of biological weapons (Sandbrink, 2023; Soice et al., 2023; Mouton, Lucas, & Guest, 2023; Gopal et al., 2023). Terrorists, cybercriminals, or other malicious actors could use such models to cause large-scale harm (Brundage et al., 2018; Anderljung & Hazell, 2023; Marchal et al., 2024). There are also increasing concerns that autonomous agents (Chan et al., 2023, 2024; Shavit et al., 2023; Cohen et al., 2024; Gabriel et al., 2024) might at some point be able to create copies of themselves, acquire resources, and seek power (Turner & Tadepalli, 2022; Turner et al., 2023; Krakovna & Kramar, 2023; Kinniment et al., 2023). Some people even worry that certain capabilities could result in a global catastrophe (Center for AI Safety, 2023; Hendrycks et al., 2023; Ngo, Chan, & Mindermann, 2022).[3] Against this background, managing the risks of frontier AI models is an important and urgent challenge.

To rise to this challenge, companies like OpenAI, Google DeepMind, and Anthropic need to strengthen their risk management practices (Bengio et al., 2024). In some cases, they need to develop new solutions, such as alignment techniques (Ji et al., 2023; Anwar et al., 2024), model evaluations for dangerous capabilities (Shevlane et al., 2023; Phuong et al., 2024), or AI safety frameworks (Anthropic, 2023a; OpenAI, 2023f; Google DeepMind, 2024a). In other cases, it would be more appropriate to learn from other industries. Instead of reinventing the wheel, frontier AI developers should apply existing best practices in risk management to an AI context.[4] One such practice is internal audit.

This article argues that frontier AI developers need an internal audit function. Section 2 sets the stage by defining key terms and reviewing relevant literature. It also describes the role of internal audit in corporate governance based on the Three Lines Model and the Combined Assurance Framework, and gives an overview of key governance challenges in frontier AI development. Section 3 advances the main argument by discussing how internal audit can address some of these challenges, while acknowledging key limitations. Section 4 concludes with a summary of the article's main contributions and suggestions for further research.

---

[2] But note that some people claim that emergent capabilities may not be a fundamental property of scaling AI models (Schaeffer, Miranda, & Koyejo, 2023). Others have pointed out that the concept is poorly defined (Rogers & Luccioni, 2023).

[3] But note that some people think that the risks are overblown (Sætra & Danaher, 2023; Nature editorial board, 2023) or at least that the current level of risk is low (Baum, 2024).

[4] For more information on risk governance, see van Asselt & Renn (2011), Lundqvist (2015), Klinke and Renn (2021).

## 2. Setting the stage

*2.1. Terminology*

Let us first define some of the key terms used in this article, namely "internal audit", "artificial intelligence (AI)", and "frontier AI", as well as related terms like "general-purpose AI (GPAI)", "foundation model", and "artificial general intelligence (AGI)".

*Internal audit.* The term "internal audit" might cause confusion because it is used in two different ways. In corporate governance, internal audit is a technical term. It refers to a specific organizational function (e.g. a team) that evaluates the adequacy and effectiveness of a company's risk management, control, and governance processes (IIA, 2024; Nagy & Cenker, 2002). This function is organizationally independent from senior management and reports directly to the board of directors, typically its audit committee. In doing so, internal audit provides independent and objective assurance about the organization's risk management practices. Internal audit professionals are typically Certified Internal Auditors (CIA), though a certificate is not required (IIA, 2023). The most senior individual with responsibilities for internal audit services is typically called Chief Audit Executive (CAE) or Head of Internal Audit (IIA, 2020).

Outside corporate governance, the term is used more loosely. It typically refers to any audit that is done internally, as opposed to a third-party audit (Raji et al., 2020; Birhane et al., 2024). The basic idea behind auditing is that an auditor (e.g. an audit firm) evaluates whether the object of the audit (e.g. an AI system or the governance of an organization) complies with predefined audit criteria (e.g. regulations or standards) (IEEE, 2008; Brundage et al., 2020; Mökander et al., 2021; Birhane et al., 2024). For example, during a financial audit, a certified accounting firm like KPMG or EY evaluates whether a company has prepared its financial statement in accordance with a recognized accounting standard like IFRS or US GAAP. Similarly, during a model audit, an organization like like METR[5] might evaluate a model against a list of realistic autonomous tasks (Kinniment et al., 2023). An audit is external if the auditor is not part of the organization that is being audited (Raji & Buolamwini, 2019; Falco et al., 2021; Raji et al., 2022; Anderljung et al., 2023b). It is internal if the auditor is part of that organization (Raji et al., 2020).[6]

---

[5] Note that METR does not see itself as an auditor, but a research organization that seeks to advance the science of model evaluations. But since METR has been commissioned by OpenAI and Anthropic to evaluate GPT-4 and Claude models (METR, 2023), I consider them an auditor for the purposes of this article.

The two ways of using the term overlap, but there are also important differences. Internal audit as a specific function has a precisely defined scope (evaluating the organization's risk management practices), purpose (providing independent and objective assurance to the board), and position within the organization (it is organizationally independent from senior management and reports directly to the board). In contrast, internal audit as the opposite of external audit is less precisely defined. It could assess models, applications, or governance structures (Mökander et al., 2023), which may or may not include the organization's risk management practices. For the purpose of this article, I use the term in the first sense: I mean the specific function that evaluates the effectiveness of risk management practices, not the opposite of external audit.

*Artificial intelligence*. There is no generally accepted definition of the term "artificial intelligence". It was first mentioned by McCarthy et al. (1955) in the funding proposal for the famous Dartmouth workshop, which is widely considered to be the birthplace of the field (Nilsson, 2009). Since then, a vast spectrum of definitions has emerged (e.g. Minsky, 1969; Kurzweil, 1990; McCarthy, 2007; Nilsson, 2009). Categorizations of different definitions have been proposed by Russell & Norvig (2021), Wang (2019), and Bhatnagar et al. (2018). However, for the purposes of this article, a precise definition is not necessary. A social definition is sufficient, according to which AI is what people generally consider to be AI (Cihon et al., 2021).[7] At the moment, most people seem to mean deep neural networks (DNNs) trained on large datasets with large amounts of compute using supervised, unsupervised, or reinforcement learning (RL) algorithms (Maslej et al., 2024). This includes large language models (LLMs) like GPT-4o (OpenAI, 2024c), Claude 3.5 Sonnet (Anthropic, 2024b, 2024c), and Gemini 1.5 (Google DeepMind, 2024b), image-generation models like DALL·E 3 (OpenAI, 2023a; Betker et al., 2023), and Stable Diffusion 3 (Esser et al., 2024; Rombach et al., 2021), and video-generation models like Sora (Brooks et al., 2024). More recently, large multimodal models (LMMs) have become popular, i.e. models that can process different types of inputs like text, image, audio, and video (Maslej et al., 2024).

*Frontier AI*. The term "frontier AI" has only recently gained traction. It has been used in several publications (e.g. Shevlane et al., 2023; Anderljung et al., 2023a; Bucknall & Trager, 2023) and in the context of the UK AI Safety

---

[6] Birhane et al. (2024) distinguish between whether or not there is a contractual relationship between the auditor and the company. According to this interpretation, METR would be considered an internal auditor.

[7] But note that this changes over time. As famously put by John McCharthy: "as soon as it works, no one calls it AI any more" (Meyer, 2011). Also note that this approach does not work in a regulatory context where a more precise definition is necessary (Schuett, 2023a).

Summit in Bletchley Park (DSIT, 2023a, 2023b, 2023c, 2023d) and the AI Seoul Summit 2024 (DSIT, 2024a, 2024b). It has also been adopted by several companies, including OpenAI (2023c, 2023e, 2023f), Google DeepMind (2023, 2024a), and Anthropic (2023a, 2023b), as well as the Frontier Model Forum (2023). In this article, "frontier AI models" are defined as highly capable general-purpose AI models that can perform a wide variety of tasks and match or exceed the capabilities present in the most advanced models (DSIT, 2024b).[8] Similar definitions have been suggested by Shevlane et al. (2023),[9] Anderljung et al. (2023),[10] and Phuong et al. (2024).[11] But note that the term has also been criticized (Helfrich, 2024). By "frontier AI developer", I mean any organization that develops frontier AI models.

*Related terms*. The term frontier AI model overlaps with several other terms. For example, a frontier model is a special type of "general-purpose AI (GPAI)" model. While the term frontier AI is defined relative to existing capabilities, the term GPAI puts more emphasis on the generality of the model's capabilities (Gutierrez et al., 2023). The term GPAI, in turn, is often used synonymously with "foundation model" (Jones, 2023), defined as any model trained on broad data that can be adapted to a wide range of downstream tasks (Bommasani et al., 2021). The main difference is that the term foundation model puts more emphasis on the model's role in the supply chain: it serves as a foundation for other models and applications. Another related term is "artificial general intelligence (AGI)", which can be defined as an AI system that achieves or exceeds human performance across a wide range of cognitive tasks (Goertzel & Pennachin, 2007; Goertzel, 2014; Altman et al., 2023; Morris et al., 2023). While some scholars think that current models at the frontier should already be considered AGI (Bubeck et al., 2023), others remain skeptical that AGI will ever be built (Fjelland, 2020; Mitchell, 2021, 2024).

---

[8] The definition could be operationalized by defining a training compute threshold, i.e. a certain amount of computational resources required to train a model (Heim & Koessler, 2024; Hooker, 2024; Koessler, Schuett, & Anderljung, 2024). A threshold suggested in the literature (Sastry et al., 2024; Anderljung et al., 2023a) and used in the recent Executive Order on Safe, Secure, and Trustworthy AI (The White House, 2023) is $10^{26}$ floating point operations (FLOP). This would be more than any existing model (Our World in Data, 2023).

[9] Shevlane et al. (2023) define "frontier AI models" as "models that are both (a) close to, or exceeding, the average capabilities of the most capable existing models, and (b) different from other models, either in terms of scale, design (e.g. different architectures or alignment techniques), or their resulting mix of capabilities and behaviors".

[10] Anderljung et al. (2023) define "frontier AI models" as "highly capable foundation models that could possess dangerous capabilities sufficient to pose severe risks to public safety".

[11] Phuong et al. (2024) define "frontier AI models" as "the leading edge of general-purpose AI models"

*2.2. Related work*

To contextualize the article within the literature, let us now review previous work on internal audit in general, internal audits of information technologies (IT), internal audits of AI, and frontier AI governance more broadly.

*Literature on internal audit*. Internal audit is a well-studied governance measure (Kotb, Elbardan, & Halabi, 2020; Roussy & Perron, 2018; Cascarino, 2015; Coetzee et al., 2024; Büchling, Cerbone, & Maroun, 2023). There are numerous empirical studies on the value of internal audit (Lenz & Hahn, 2015; Eulerich & Eulerich, 2020; Jiang, Messier, & Wood, 2020). For example, internal audit is associated with increased financial performance (Jiang, Messier, & Wood, 2020), a decline in perceived risk (Carcello et al., 2020), strengthened internal controls (Lin et al., 2011; Oussii & Taktak, 2018), and improved fraud prevention (Coram, Ferguson, & Moroney, 2008; Ma'ayan & Carmeli 2016; Drogalas et al., 2017). There is also literature on the drivers of internal audit effectiveness (Arena & Azzone, 2009; Soh & Martinov-Bennie, 2011; Coetzee & Lubbe, 2014; Erasmus & Coetzee, 2019), the challenge of becoming or remaining independent from senior management (Stewart & Subramaniam, 2010; Roussy, 2013; Guénin-Paracini, Malsch, & Tremblay, 2015; Roussy & Rodrigue, 2018; Nordin, 2023), and the future of internal audit (Chambers & Odar, 2015; Christ et al., 2021; Betti & Sarens, 2021).

*Literature on internal IT audits*. There is also extensive literature on internal IT audits, i.e. internal audits that assess the effectiveness, efficiency, and security of IT systems and processes (Hermanson, Hill, & Ivancevich, 2000; Weidenmier & Ramamoorti, 2006; Merhout & Havelka, 2008; Senft & Gallegos, 2009). There are numerous empirical studies that examine the value and effectiveness of internal IT audits as part of IT governance (De Haes & Van Grembergen, 2009). For example, Steinbart et al. (2012) find that internal IT audits can improve information security, while Stafford et al. (2018) investigate the role of IT audits in identifying non-compliance with security policies in the workplace. Stoel et al. (2012) identify factors that contribute to IT audit quality, including the technical competence of auditors and the quality of client relationships, while Merhout and Havelka (2008) identify six critical factors for audit success and argue for a value-added partnership between IT management and auditors.

*Literature on internal audit and AI*. Internal AI audits can be seen as a subset of internal IT audits (ISACA, 2018). Although there is some literature on the intersection of internal audit and AI, most of it is about how internal auditors can use AI (Couceiro, Pedrosa, & Marini, 2020; Kahyaoglu & Aksoy, 2021; Emett et al., 2023; Wassie & Lakatos, 2024).[12] The most relevant piece is an article that applies the Three Lines Model, a risk governance framework where internal audit serves as the third line, to an AI context (Schuett, 2023c). Here,

---

[12] Note that I excluded literature that follows the second interpretation of internal audit (Section 2.1), such as Raji et al. (2020) and Birhane et al. (2024).

I conduct a more in-depth analysis of the third line. Besides that, the Institute of Internal Auditors (IIA) has published a three-part series, in which they propose an AI auditing framework (IIA, 2017a, 2017b, 2018). In the first two parts, they specifically discuss the role of internal audit, though the relevant passages are rather short and somewhat outdated. There is only one study that mentions internal audit in the context of AGI developers (Schuett et al., 2023), but only as part of a broader expert survey on best practices in AGI safety and governance. There does not seem to be an investigation of the benefits and limitations of internal audit at frontier AI developers.

*Literature on frontier AI governance*. The study of frontier AI governance is a small but growing field. Ahead of the UK AI Safety Summit, the UK Department for Science, Innovation and Technology (DSIT) published a policy paper on emerging processes in frontier AI safety (DSIT, 2023c). At the AI Seoul Summit 2024, 16 companies agreed to the Frontier AI Safety Commitments (DSIT, 2024b). Besides that, there is literature on frontier AI regulation (Anderljung et al., 2023a; Schuett et al., 2024), evaluating frontier models for dangerous capabilities (Shevlane et al., 2023; Phuong et al., 2024), setting risk thresholds for frontier AI (Koessler, Schuett, & Anderljung, 2024), responsible reporting for frontier AI development (Kolt et al., 2024), responding to risks that are discovered after a frontier model has been deployed (O'Brien, Ee, & Williams, 2023), promoting external scrutiny of frontier models (Bucknall & Trager, 2023; Anderljung et al., 2023b), overseeing frontier AI through know-your-customer (KYC) requirements for compute providers (Egan & Heim, 2023; Sastry et al., 2024), and establishing an international governance regime for frontier AI (Ho et al., 2023; Gruetzemacher et al., 2023; Trager et al., 2023). It is worth noting, however, that not all relevant literature uses the term frontier AI. Some of it is about GPAIS (e.g. Barrett et al., 2023), foundation models (e.g. Partnership on AI, 2023; Seger et al., 2023b; Vipra & Korinek, 2023; Kapoor et al., 2024), generative AI (e.g. Hacker, Engel, & Mauer, 2023; Longpre et al., 2024), or AGI (e.g. Schuett et al., 2023; Koessler & Schuett, 2023).

### 2.3. The role of internal audit in corporate governance

Internal audit plays a key role in corporate governance. Below, I describe its role based on the Three Lines Model and the Combined Assurance Framework. This provides the theoretical foundation for the main argument in Section 3.

*Three Lines Model*. Risk management roles and responsibilities are increasingly split across multiple teams (e.g. legal, compliance, and cybersecurity). However, without proper coordination, work can be duplicated and gaps in risk coverage can occur (Bantleon et al., 2021). The Three Lines Model, formerly known as Three Lines of Defense (IIA, 2013; Davies & Zhivitskaya, 2018), is intended to address this problem (IIA, 2020). It is a popular risk governance framework that helps organizations to assign and coordinate risk management roles and responsibilities. It distinguishes between three roles, which it calls

"lines". The first line provides products and services to clients and is ultimately responsible for risk management. The second line assists the first line. It provides complementary expertise and support, but also monitors and challenges the first line. The third line provides independent and objective assurance and advice to the governing body (i.e. the board of directors). It is not directly involved in core operations. External assurance providers such as third-party auditors and supervisory authorities provide additional assurance. They are even more independent than internal audit. The model is illustrated in Figure 1. Although it has been criticized (Arndorfer & Minto, 2015; Leech & Hanlon, 2016), it remains "the most carefully articulated risk management system that has so far been developed" (Davies & Zhivitskaya, 2018). Most listed companies have implemented the model; it is particularly popular in the financial industry (Huibers, 2015; Bantleon et al., 2021).

In the Three Lines Model, internal audit serves as the third line. It is separate from the first two lines and reports directly to the board of directors, typically its audit committee (Sarens, De Beelde, & Everaert, 2009). The board plays a key role in corporate governance. It sets the company's strategic priorities, is responsible for risk oversight, and has significant influence over management (e.g. it can replace senior executives) (Zald, 1969). However, since non-executive board members only work part-time, they rely on information provided to them by the executives (Davies & Zhivitskaya, 2018), which might only tell the board what they think it wants to hear, not what it needs to hear. As a result, it can be difficult for the board to get an accurate view of the current level of risk and the effectiveness of the company's risk management practices. This situation can be phrased as a principal agent-problem: there is an information asymmetry (problem) between the board, which is legally responsible for risk oversight (principal), and the executives, who are responsible for risk-related activities (agents). The role of internal audit is to tackle this problem by providing the board with independent and objective information. It is often described as the board's "eyes and ears" (IIA, 2020). While the Chief Risk Officer (CRO), the most senior executive responsible for risk management (Karanja & Rosso, 2017; Li et al., 2022), reports on the current level of risk and outcomes of risk management activities, the Chief Audit Executive (CAE) tells the board how much they can trust these reports (e.g. "their method for evaluating risks is flawed"). Since internal audit is organizationally independent from senior management, it is less biased and more objective. With these two reporting lines, the board has a more complete picture.

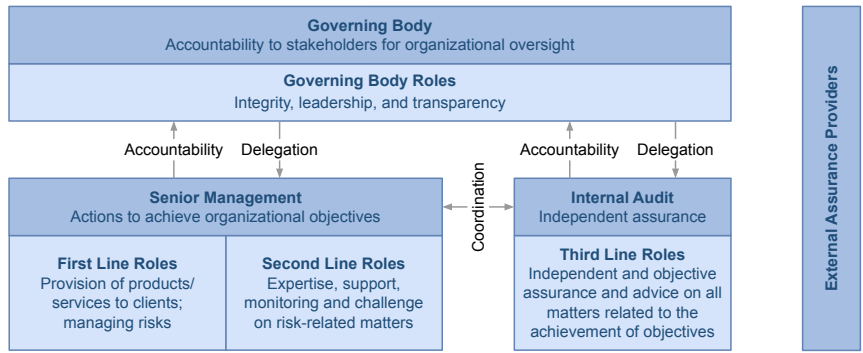

*Figure 1*: IIA's Three Lines Model

*Combined Assurance Framework*. But internal audit is not the only assurance provider. In many companies, the board also gets reports from other internal assurance providers (e.g. compliance or quality control) and external assurance providers (e.g. external auditors), which might have radically different perspectives on assurance (Roussy & Brivot, 2016). If the different assurance activities are performed in isolation, the board can suffer from "assurance fatigue" and assurance gaps can occur (Sarens, Decaux, & Lenz, 2012). Put simply, more assurance is not always better. Against this background, the purpose of Combined Assurance is to "provide holistic assurance to the board on the effectiveness of risk management and internal control systems by coordinating assurance activities from various sources of assurance" (Decaux & Sarens, 2015).[13]

Combined Assurance has a number of advantages, including (1) improved knowledge transfer between internal parties, (2) more efficient use of resources by eliminating overlapping efforts, (3) focus on higher risk, and (4) improved communication with senior management and the board (IIA, 2022). However, many companies seem to find it challenging to implement the framework (PwC, 2015). A common obstacle is that the Three Lines Model is implemented poorly. Blurred lines might result in irrelevant, inefficient, or inadequate coordination of assurance efforts. Against this background, a clear allocation of assurance responsibilities and communication between the lines is crucial (Huibers, 2015; PwC, 2015; IIA, 2020, 2022).

Since internal audit is typically the main internal assurance provider, it has a special responsibility for promoting Combined Assurance (Huibers, 2015). This is also reflected in Standard 9.5 of the IIA's Global Internal Audit Standards, which states that "the chief audit executive must coordinate with internal and external providers of assurance services and consider relying upon their work. Coordination of services minimizes duplication of efforts, highlights gaps in coverage of key risks, and enhances the overall value added by providers" (IIA, 2024). Figure 2 illustrates Combined Assurance within the Three Lines Model.

---

[13] For more information on the Combined Assurance Framework, see PwC (2015), Huibers (2015), Forte and Barac (2016), Prinsloo and Maroun (2021), IIA (2022).

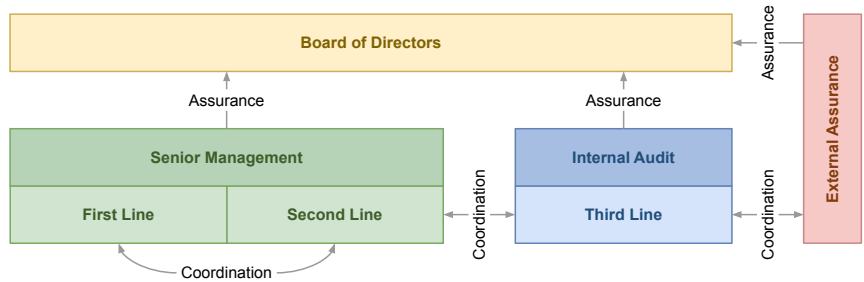

*Figure 2*: Combined Assurance within the Three Lines Model

Before we can discuss whether frontier AI developers need an internal audit function, we first need to understand the specific governance challenges they face.

### 2.4. Key governance challenges in frontier AI development

The governance of frontier AI development raises many challenges—from ensuring broad participation in high-stakes governance decisions (Birhane et al., 2022a; Delgado et al., 2023; Seger et al., 2023a) to managing labor market impacts (Frank et al., 2019; Eloundou et al., 2024). However, this article focuses on governance challenges that may arise in the context of reducing societal risks from frontier AI models.[14] Below, I list five key challenges.[15] They are summarized in Table 1.

---

[14] Although this article focuses on risks, internal audit could also help to address other ethical challenges. For example, many AI companies have ethics principles (Jobin, Ienca, & Vayena, 2019; Hagendorff, 2020), but "principles alone cannot guarantee ethical AI" (Mittelstadt, 2019). Internal audit could evaluate the extent to which principles are put into practice (Morley et al., 2020, 2023). It could also advise the board on ethical matters. For example, it could ensure that developers pay enough attention to their impact on historically marginalized communities (Mohamed, Png, & Isaac, 2020; Birhane et al., 2022a, 2022b), the environment (van Wynsberghe, 2021), and animals (Singer & Tse, 2023).

[15] For more information on the first three governance challenges, see Anderljung et al. (2023). Note that the five challenges are neither mutually exclusive nor collectively exhaustive. They were selected because they seem particularly relevant for the purposes of this article.

| Key challenges | Description |
| --- | --- |
| The unexpected capabilities problem | Dangerous capabilities can arise unpredictably and undetected, both during development and after deployment. |
| The deployment safety problem | It is difficult to prevent a deployed model from causing harm. |
| The proliferation problem | Frontier models can proliferate rapidly. |
| The risk assessment problem | It is inherently difficult to assess frontier AI risks. |
| The risk governance problem | Frontier AI developers do not follow best practices in risk governance. |

*Table 1*: Key governance challenges in frontier AI development

*The unexpected capabilities problem*. Dangerous capabilities can arise unpredictably and undetected, both during development and after deployment. In general, model performance tends to improve smoothly with more data, more parameters, and more compute (Kaplan et al., 2020; Bahri et al., 2021; Hoffmann et al., 2022). However, specific capabilities can emerge sharply, transitioning seemingly instantaneously from not present to present, and unpredictability, appearing at seemingly unforeseeable model scales (Ganguli et al., 2022a; Wei et al., 2022).[16] But as mentioned above, some of these capabilities might be dangerous (Shevlane et al., 2023; Phuong et al., 2024). Although there is increasing interest in model evaluations for dangerous capabilities (Shevlane et al., 2023; Phuong et al., 2024), the field is still nascent and it is not feasible to test for all relevant capabilities (Burnell et al., 2023; Burden, 2024). As a result, some capabilities are discovered long after a model has been deployed. This phenomenon has been called "capability overhang" (Clark, 2022). It is also possible to actively enhance certain dangerous capabilities. For example, malicious actors might fine-tune the model on a task-specific dataset (Goldstein et al., 2023) or combine it with external tools (Schick et al., 2023; Cai et al., 2023; Mialon et al., 2023; Davidson et al., 2023).

*The deployment safety problem*. It is difficult to prevent a deployed model from causing harm. One reason is that reliably controlling the behavior of frontier models, also known as "alignment", remains an unsolved technical problem (Amodei et al., 2016; Gabriel, 2020; Kenton et al., 2021; Hendrycks et al., 2021; Ngo, Chan, & Mindermann, 2022; Ji et al., 2023; Anwar et al., 2024).[17]

---

[16] For a more skeptical perspective, see Schaeffer, Miranda, & Koyejo (2023).

[17] It is worth noting that techniques like reinforcement learning from human feedback (RLHF) (Christiano et al., 2017; Ziegler et al., 2019; Lampert et al., 2022) and reinforcement learning from AI feedback (RLAIF), more commonly known as "constitutional AI" (Bai et al., 2022), represent major advancements.

Another reason is that safety measures can be circumvented. For example, it is surprisingly easy to "jailbreak" a model (Yong, Menghini, & Bach, 2023; Anil et al., 2024) or to bypass its safety filters (Rando et al., 2022). It is also possible to remove safety measures like reinforcement learning from human feedback (RLHF) by fine-tuning the model (Zhan et al., 2023; Lermen, Rogers-Smith, & Ladish, 2023; Gade et al., 2023). Finally, frontier models are dual use: they can be used for both beneficial and harmful purposes (Urbina et al., 2022; Anderljung & Hazell, 2023). For example, language models can be used to write professional emails, but also phishing messages (Hazell, 2023). Since the harmfulness of frontier models is often highly contextual (Weidinger et al., 2023), not merely a model property (Narayanan & Kapoor, 2024), mitigating misuse risks is particularly challenging (Clifford, 2023).

*The proliferation problem.* Frontier models can proliferate rapidly. There are at least four ways in which this can happen. First, many developers make their models available via an application programming interface (API) (Shevlane, 2022; Solaiman, 2023; Bucknall & Trager, 2023). This includes models like GPT-4 (OpenAI, 2023b), Gemini (Google DeepMind, 2023b), and Claude 3 (Anthropic, 2024b). Although they typically restrict who can access the model and how they can use it (Soleiman et al., 2019; O'Brien, Ee, & Williams, 2023), it remains difficult to reliably prevent cases of misuse (see above). Second, models might get reproduced. Popular models often get reproduced or improved upon within 1-2 years of their initial release (Zhao et al., 2023). The developers of the reproduced models are typically independent researchers who do not have the resources to take extensive safety measures. Third, some developers open-source their models, i.e. they make the model architecture and weights freely and publicly accessible.[18] Llama 3 (Meta, 2024) or BLOOM (Scao et al., 2022) are two popular examples. Since open-source models provide significant societal benefits, many people in the AI community are generally pro open source (Shrestha, von Krogh, & Feuerriegel, 2023). However, at some point—though arguably not yet (Kapoor et al., 2024)—the risks may outweigh the benefits (Seger et al., 2023b; Bateman et al., 2024). But as mentioned above, it is difficult to know in advance if a model has certain dangerous capabilities, and once a model has been open-sourced, it cannot be taken back. Fourth, malicious actors might steal the model weights—potentially even before the model has been released. Although some developers take extensive measures to secure their models (Anthropic, 2023b), sufficient defenses are becoming increasingly costly (Nevo et al., 2024).

*The risk assessment problem.* It is inherently difficult to assess risks from frontier AI models. One reason is that many risks are unprecedented. Although a wide range of potential risks have already been identified (Weidinger et al., 2021; Bommasani et al., 2021; Hendrycks et al., 2023; Slattery et al., 2024), some risks will most likely be missing, including so-called "black swans" (Taleb, 2007; Aven, 2013, 2016; Kolt, 2023). The unprecedented nature of many risks also makes it difficult to estimate their impact and likelihood. Since historical data is limited, developers often have to rely on forecasts. Although there are robust ways to improve forecasting accuracy (Tetlock, 2005; Tetlock & Gardner, 2015; Chang et al., 2016; Petropoulos et al., 2022), estimates will be more uncertain and less reliable. Estimates would be even less reliable if, at some point, models become able to detect when they are being tested and only follow instructions in a testing environment (Carlsmith, 2023; Carranza et al., 2023; Berglund et al., 2023). Another reason why assessing frontier AI risk is difficult is that the field is still in its infancy. Despite encouraging developments in some domains, most notably model evaluations (Liang et al., 2021; Srivastava et al., 2022; Shevlane et al., 2023; Kinniment et al., 2023; Phuong et al., 2024; Weidinger et al., 2024), best practices have not yet emerged (Schuett et al., 2023). Compared to other safety-critical industries like nuclear and aviation, current methods in frontier AI risk assessment seem much less sophisticated (Koessler & Schuett, 2023). Against this background, we should expect that some risk assessment practices will be flawed and not fit-for-purpose (Rae, Alexander, & McDermid, 2014; Røyksund & Flage, 2018).

---

[18] For an overview of different deployment modes, see Solaiman (2023).

*The risk governance problem.* Although, some developers have implemented innovative governance structures—for example, the OpenAI nonprofit governs the OpenAI LP and owns all returns above a certain threshold (OpenAI, 2019), while Anthropic's Long-Term Benefit Trust (LTBT) has the power to elect and remove some of Anthropic's board members (Anthropic, 2023c)—they mostly do not follow existing best practices in risk governance.[19] In particular, they do not seem to have established a board risk committee, appointed a chief risk officer (CRO), set up an internal audit function, or implemented the Three Lines Model. It is unclear if they even have a central risk function. For example, OpenAI's Preparedness team is only responsible for managing catastrophic risks from frontier AI (OpenAI, 2023c), while Google DeepMind's Responsibility and Safety Council (RSC) is mostly concerned with upholding their AI ethics principles (Google DeepMind, 2023b). They do not seem to have a team that is responsible for holistic risk management, which is a key element of modern enterprise risk management (ERM) (Bromiley et al., 2015; McShane, 2018). It remains to be seen if OpenAI's recent announcement to create a cross-functional Safety Advisory Group (SAG) will serve this function (OpenAI, 2023f). Finally, frontier AI developers do not seem to have a function that is responsible for evaluating the effectiveness of their risk management practices. This is a major shortcoming in light of the risk assessment problem mentioned above.

---

[19] Note that the following is based on public information. It is possible that developers have not made certain structures public or use different terms.

## 3. Advancing the argument

In this section, I argue that frontier AI developers need an internal audit function. While internal audit already plays a key role in the governance of information technology (ISO/IEC, 2024), the following discussion focuses on how internal audit can address some of the specific governance challenges related to the development of frontier AI models, while also acknowledging key limitations.

*3.1. How internal audit can address some of the governance challenges in frontier AI development*

There are at least three ways in which internal audit can address some of the above-mentioned governance challenges.

*Identifying ineffective risk management practices.* First, internal audit can help to identify ineffective or inadequate risk management practices. For example, it could assess the accuracy and reliability of the developer's model evaluations (Liang et al., 2021; Srivastava et al., 2022; Shevlane et al., 2023; Kinniment et al., 2023; Phuong et al., 2024; Weidinger et al., 2024). There are at least four reasons why their evaluations might be inaccurate or unreliable: (1) Developers might simply not conduct evaluations of certain capabilities. To avoid blind spots, internal audit could scrutinize the developer's process for deciding what evaluations to run, and engage with industry bodies like the Frontier Model Forum (FMF) to learn more about what evaluations other developers run. (2) It is often unclear how much information evaluations in a lab environment provide about a model's behavior in the real world. To better understand how evaluation results generalize, internal audit could attempt to predict future evaluation results based on an initial set of evaluations (Chan, 2024). This would help decision-makers to calibrate their trust in evaluation results. (3) Evaluations are often conducted by the same people who develop the model.[20] Since researchers and engineers have strong incentives to quickly develop and deploy models, they might—consciously or not—downplay or ignore concerning evaluation results. To reduce this risk, internal audit could oversee discussions of evaluation results and challenge potentially biased interpretations. (4) Most evaluations only look at a model's behavior, without understanding the reasons for the behavior (Hubinger, 2023).[21] This approach

---

[20] Note that this refers to evaluations that are conducted at various checkpoints during the training process. Before deploying a new model, a separate team might also conduct "assurance evaluations", using prompts unavailable to the development team (Weidinger et al., 2024). Some developers also commission external model evaluations. For example, METR evaluated both GPT-4 and Claude models (METR, 2023).

[21] This is a result of the black box nature of current AI models (Castelvecchi, 2016; Rudin, 2019; von Eschenbach, 2021). But note that there is an emerging research field called

would be unreliable if, at some point, models become able to deceive the people who conduct the evaluations and only pretend to behave in a certain way (Park et al., 2023; Hagendorff, 2023; Carlsmith, 2023; Pacchiardi et al., 2023; Hubinger et al., 2024; Anthropic, 2024a; Järviniemi & Hubinger, 2024), a concern that has been coined "sandbagging" (van der Weij et al., 2024). Depending on their technical expertise, internal auditors could empirically test the extent to which a model demonstrates deceptive capabilities and tendencies, perhaps similar to Anthropic's Alignment Stress-Testing Team (Hubinger, 2024; Hubinger et al., 2024; MacDiarmid et al., 2024).

Internal audit could also assess the adequacy of the developer's security measures (Anthropic, 2023b). For example, a recent RAND report suggests that frontier AI developers are currently unable to defend against cyberattacks from advanced threat actors (Nevo et al., 2024). Internal audit could commission an external red team to identify vulnerabilities. It could also monitor compliance with relevant security standards such as the NIST Secure Software Development Framework (SSDF) (NIST, 2022) and the Supply Chain Levels for Software Artifacts (SLSA) (OpenSSF, 2023).

In addition to that, internal audit could verify whether the developer complies with its AI safety framework.[22] For example, Anthropic's Responsible Scaling Policy (RSP) (Anthropic, 2023a), OpenAI's Preparedness Framework (OpenAI, 2023f), and Google DeepMind's Frontier Safety Framework (Google DeepMind, 2024a) all contain commitments to pause the development and deployment process if their safety measures are inadequate for the model's level of capabilities. But since a pause would be extremely costly for the developer, the people involved have strong incentives to conclude that the safety measures are adequate (Alaga & Schuett, 2023). They might adopt a box-ticking mindset and only superficially comply with the commitments in their framework. To find out if this is the case, internal audit could attend meetings, conduct interviews, and review documents. Taken together, these examples illustrate that, without a deliberate attempt to identify ineffective or inadequate risk management practices, some shortcomings will likely remain unnoticed.

*Better informed board of directors.* Second, internal audit can ensure that the board of directors has a more accurate view of the current level of risk and the effectiveness of the company's risk management practices. At many frontier AI developers, the board of directors plays a special role. The recent scandal around some of OpenAI's board members—who first fired CEO Sam Altman (OpenAI, 2023d), and then had to resign after internal criticisms and a public outcry (OpenAI, 2023g, 2024b)—serves as a cautionary example. The board members reportedly felt that Altman withheld important information and even lied to them (Duhigg, 2023). However, since many details of the case are

---

"mechanistic interpretability" that tries to understand how models work by studying their weights and activations (Olah et al., 2020; Nanda et al., 2023).

[22] For more information on verifying claims about responsible AI development, see Brundage et al. (2020) and Avin et al. (2021).

not yet public, such reports should be taken with a grain of salt. It is difficult to tell what, if anything, an internal audit function would have changed in this particular case. But it shows that information asymmetries between the board and senior management can be extremely consequential. Shortly after the board scandal, OpenAI published their Preparedness Framework ([OpenAI, 2023f](#)). According to this framework, the new board needs to be involved in certain decisions and can even overrule senior management. Anthropic's RSP contains similar provisions ([Anthropic, 2023a](#)). For example, the board needs to approve updates of the RSP. As the boards of frontier developers are increasingly involved in high-stakes decisions, they need independent and objective information about those decisions. Setting up an internal audit function could help to address this challenge.

*Contact point for whistleblowers*. Third, internal audit could serve as a contact for whistleblowers. Detecting misconduct is often difficult: it is hard to observe from the outside, while insiders might not report it because they face a conflict between personal values and loyalty ([Jubb, 1999](#); [Dungan, Waytz, & Young, 2015](#)), or because they fear retaliation ([Bjørkelo, 2013](#)). For example, an engineer might become convinced that a model shows early signs of power-seeking behavior ([Ngo, Chan, & Mindermann, 2022](#); [Turner & Tadepalli, 2022](#); [Carlsmith, 2022](#); [Turner et al., 2023](#); [Krakovna & Kramar, 2023](#)), but the research lead might want to release the model anyway and threatens to fire the engineer if they speak up. In such cases, whistleblower protection is vital. This might become even more important as commercial pressure increases because it might incentivize developers to cut corners on safety ([Armstrong, Bostrom, & Shulman, 2016](#); [Naudé & Dimitri, 2020](#)). Internal audit could protect whistleblowers by providing a trusted contact point. It would be more trustworthy than other organizational units because it is organizational independent from management. But since it would still be part of the organization, confidentiality would be less of a problem. This can be particularly important if the information is highly sensitive and its dissemination could be harmful in itself ([Bostrom, 2011](#); [Urbina et al., 2022](#)). Internal audit could report the case to the board of directors who could engage with management to address the issue. In the example, this could avoid rapid proliferation of a model with potentially dangerous capabilities. Internal audit could also advise the whistleblower on other steps they could take to protect themselves or do something about the misconduct. Finally, it increases the chance that concerns during the development phase are also taken into consideration. While this is not a typical role of internal audit, it has been discussed in the literature ([Jubb, 1999](#)).

### 3.2. Limitations

However, internal audit also has limitations. The following limitations seem particularly relevant for frontier AI developers, even though they are not unique to AI.

*Internal audit adds friction.* The internal audit team interacts with many different people, including C-suite executives and senior researchers. To them, internal audit might seem annoying, distracting, and bureaucratic. They might even be actively opposed if they fear that internal audit discovers flaws in their work,[23] especially if they can be held personally liable for their mistakes ([Thekdi & Aven, 2022](#)). Internal audit might also (indirectly) delay decisions. For example, OpenAI spent six months on safety research, risk assessment, and iteration before releasing GPT-4 ([OpenAI, 2023b](#)). While this level of scrutiny is commendable, it seems plausible that internal audit would have found flaws in some of their risk assessment methods. Depending on the severity of these flaws, internal audit would presumably have escalated the issue to the board, which might have started an investigation that might have resulted in an improvement of the methods and a repetition of the initial assessment. This could have delayed the release for additional weeks or months. To be clear, such a delay would not only be in the interest of society, it might also be in the developers' own interest. Rushed releases can expose companies to significant financial and reputational risks. For example, Alphabet's share price dropped 9%—$100 billion in market value—after Google's chatbot Bard made some mistakes in a public demo ([Coulter & Bensinger, 2023](#)). Similarly, Microsoft faced negative press coverage after users reported that Bing Chat sometimes gives bizarre answers and even threatens users ([Roose, 2023](#); [Perrigo, 2023](#)).

*Internal audit can be captured by senior management.* In theory, the internal audit function should be organizationally independent from senior management. In practice, however, senior management often finds informal ways to exercise influence. Becoming and remaining independent is an ongoing challenge of every internal audit function ([Stewart & Subramaniam, 2010](#); [Roussy, 2013](#); [Guénin-Paracini, Malsch, & Tremblay, 2015](#); [Roussy & Rodrigue, 2018](#); [Nordin, 2023](#)). Against this background, external assurance providers are an important complement. In the context of frontier AI developers, this mainly includes external auditors ([Raji & Buolamwini, 2019](#); [Raji et al., 2022](#); [Mökander & Floridi, 2022](#); [Mökander et al., 2023](#); [Anderljung et al., 2023b](#); [Birhane et al., 2024](#)), independent red teams ([Ganguli et al., 2022b](#); [Perez et al., 2022](#); [Anthropic, 2023d](#)), and ethics boards ([Schuett, Reuel, & Carlier, 2023](#)).

*Benefits depend on individuals.* Simply having an internal audit function is not sufficient to seize the above-mentioned benefits. The value of internal audit mainly depends on the people involved and their ability and willingness to identify ineffective risk management practices. Unfortunately, there do not seem to be many candidates who have both AI and internal audit expertise. Since internal audit has traditionally had a strong financial focus ([Abbott et al., 2016](#)) and frontier AI is a more recent phenomenon, skill gaps and the absence of clearly developed assurance methodologies will be a challenge. Frontier AI developers could either hire internal audit professionals and train them in AI ("bring AI to internal audit"), or they could train AI governance experts in internal audit ("bring internal audit to AI"). Since both approaches have advantages and disadvantages, developers should arguably do a mix of both. The abilities of internal auditors can also be enhanced by AI. AI is already used to support internal audit in other companies (Couceiro, Pedrosa, & Marini, 2020; Kahyaoglu & Aksoy, 2021; Emett et al., 2023) and using AI to oversee AI is a common theme in the technical safety debate (Christiano, Shlegeris, & Amodei, 2018; Leike et al., 2018; Irving, Christiano, & Amodei, 2018; Bowman et al., 2022; Burns et al., 2023; Kenton et al., 2024).

---

[23] But note that Anthropic has recently set up a new team which has the explicit mandate to empirically demonstrate ways in which its alignment strategies could fail ([Hubinger, 2024](#)).

*3.3. Discussion*

The above-mentioned limitations do not provide an argument against internal audit *per se*. In particular, they are not enough to outweigh the benefits. Instead, decision-makers should be aware of the limitations and proactively address them. Below, I discuss how best practices from internal IT audits could be applied to AI and how internal AI audits can be aligned with other assurance activities. I also argue that we will eventually need frontier AI regulation.

*Internal AI audits should incorporate best practices from internal IT audits where appropriate.* Since internal AI audits can be seen as a subset of internal IT audits (ISACA, 2018), it generally makes sense to apply existing best practices from internal IT audits to AI. However, due to the specific governance challenges in frontier AI development (see Section 2.4), existing best practices from internal IT audit need to be modified and adapted to an AI context. Below, I summarize six critical factors that influence internal IT audit quality, as identified by Merhout and Havelka (2008), discuss how they can be applied to internal AI audits, and flag potential challenges (Table 2).

| Best practices in internal IT audit (Merhout & Havelka, 2008) | Applying best practices from internal IT audits to internal AI audits | Potential challenges of applying best practices from internal IT audits to internal AI audits |
|---|---|---|
| Internal IT audit teams should use an audit methodology. | Internal AI audit teams should use an AI-specific audit methodology informed by best practices in frontier AI safety and governance. | • Internal audit methodologies for frontier AI do not yet exist.<br>• Due to the specific governance challenges of frontier AI, existing IT audit methodologies are insufficient for frontier AI.<br>• Creating audit methodologies for frontier AI is difficult because best practices in frontier AI safety and governance do not yet exist.<br>• Audit methodologies for frontier AI might quickly become outdated because the AI risk landscape and the state of the art in AI safety and governance are rapidly evolving. |
| Internal IT audit teams should have enough time to conduct audits (e.g. for fieldwork). | Internal AI audit teams should have enough time to deeply engage with and scrutinize the developer's safety and governance practices (e.g. threat models, model evaluations, and alignment techniques). | • Frontier AI developers have strong economic incentives to release models quickly. Internal audit teams might therefore be pressured to rush audits. |
| Internal IT audit teams should have support from senior management. | Internal AI audit teams should have support from senior management, especially the executives who are responsible for AI research, product development, and risk management. | • At least historically, frontier AI developers wanted to "move fast and break things". They were generally opposed to formal processes and structures.<br>• Senior management might not see the need to support internal audit. This is because frontier AI developers are not legally required to set up an internal audit function. It is also unclear to what extent key stakeholders like business partners and investors expect developers to set up an internal audit function on a voluntary basis. |
| Auditees should respond quickly to inquiries from the internal IT audit team. | Teams working on frontier AI safety and governance (e.g. members of the alignment team) | • At some frontier AI developers, there seems to be a divide between product and safety teams, which makes it more challenging to |
| Communication should be honest and open. The relationship should be based on mutual trust. | should respond quickly to inquiries from the internal AI audit team. Developers should promote collaboration and trust between different teams. | promote collaboration and trust.<br>• Trust builds over time. But since developers currently do not have an internal audit function, we should expect some initial reservations. |
| Internal IT audit teams should have support to adapt to organizational change (e.g. restructurings of business processes). | Internal AI audit teams should have support to adapt to organizational change (e.g. by hiring more auditors and giving them more time). | • It might be difficult to get the necessary support given the potential lack of buy-in from senior management and the divide between product and safety teams (see above). |
| The scope and objectives of IT audits should be clearly defined. | The scope and objectives of the AI audit should be clearly defined. | • The definition of the scope and objectives may need to be updated frequently because the AI risk landscape and the state of the art in AI safety and governance is rapidly evolving. |

*Table 2*: Applying best practices from internal IT audits to internal AI audits

*Internal audit needs to be aligned with other assurance providers.* Aligning different assurance providers is particularly important in the context of frontier AI. Due to the specific governance challenges of frontier AI development (see Section 2.4), it is important that senior management and the board of directors have a good understanding of the level of risk and the adequacy of risk management practices, including key uncertainties and potential deficiencies. They need multiple sources of assurance. However, if these sources are not aligned, senior management and the board might form incorrect views about risks or risk management practices. For example, the board of directors might think that the company's safety framework is effective, without knowing that the company has not implemented parts of the framework.

However, aligning different assurance providers of frontier AI developers raises a distinct set of challenges: (1) As of today, there is not much to align. Existing assurance activities focus mainly on red teaming (Ganguli et al., 2022b; Perez et al., 2022; Anthropic, 2023d) and model evaluations (Shevlane et al., 2023; Anthropic, 2023e; Phuong et al., 2024). There is hardly any assurance of developers' internal governance. (2) Best practices in frontier AI assurance are still emerging. It is difficult to align assurance activities if they are constantly changing. (3) Safety practices and governance structures are also changing. It is difficult to develop best practices for assurance activities if their target is constantly moving. (4) Frontier AI developers do not seem to be aware of the need to align different assurance activities. The issue has not been raised by developers, policymakers, or scholars.

To overcome these and other challenges, frontier AI developers may take the following measures: (1) Developers should expand their assurance activities. In addition to setting up an internal audit function, they should commission governance audits that assess the adequacy of their safety frameworks and verify that they adhere to them. (2) Developers should encourage different assurance providers to speak the same language. Among other things, they should use the same terms, concepts, risk taxonomies, threat models, and thresholds. (3) Developers should encourage different assurance providers to submit integrated reports where appropriate (Kolt et al., 2024). It should be avoided that each assurance provider submits their own report. (4) Developers should facilitate information sharing between different assurance providers. For example, third parties that conduct model evaluations (e.g. METR, Apollo Research, and the UK AI Safety Institute), should exchange information about evaluation protocols, results, and challenges. (5) Developers should support research on assurance methodologies for frontier AI. However, it is important that such research remains independent and is not influenced by industry interests.

*The need for frontier AI regulation*. Although it may be in developers' own interest to set up an internal audit function, we should not rely on voluntary adoption. Developers should not be able to "grade their own homework". We will eventually need frontier AI regulation (Anderljung et al., 2023a; Schuett et al., 2024), which may also include the requirement to have an internal audit function,[24] similar to other industries (e.g. European Banking Authority, 2021).[25] The EU AI Act does not contain any rules on internal audit (see Mahler, 2022; Schuett, 2023b; Novelli et al., 2023), though the Code of Practice that will concretize the requirements on GPAI model providers may do (European Commission, 2024). It remains to be seen if upcoming AI regulation in the UK will be any different (DSIT, 2022; 2023b). Although the NIST AI Risk Management Framework does not mention internal audit either, it seems to describe its role in the Govern 1.5 function: "ongoing monitoring and periodic review of the risk management process and its outcomes are planned and organizational roles and responsibilities clearly defined, including determining the frequency of periodic review" (NIST, 2023). Despite the standard's voluntary nature, it might become the basis for future AI regulation, similar to the NIST Cybersecurity Standards (NIST, 2018). The recent Executive Order on Safe, Secure, and Trustworthy AI, in which NIST plays a key role, is a promising sign in this direction (The White House, 2023).

---

[24] For a critical perspective on mandating risk governance best practices, see Enriques and Zetzsche (2013).

[25] In other industries, the requirement to have an internal audit function is often described as an abstract principle, not a prescriptive rule. This principle is then specified at the sub-regulatory level. For more information on how prescriptive frontier AI regulations should be, see Schuett et al. (forthcoming).

## 4. Conclusion

*Main contributions*. This article has made three main contributions. First, it has introduced internal audit to the frontier AI governance discourse. Although internal audit is a well-established governance mechanism in many other industries, it has been largely neglected in the debate around the governance of frontier AI development. Inversely, the article has introduced the key governance challenges of frontier AI development to the field of risk science (Aven, 2018). Second, the article has advanced the argument that frontier AI developers need an internal audit function. It has discussed how internal audit could address some of the key governance challenges in frontier AI development, while acknowledging key limitations. It has also suggested ways in which best practices from internal IT audits can be applied to internal AI audits and how internal AI audits can be aligned with other assurance activities. Third, the article contributes to the creation of best practices in frontier AI governance (Schuett et al., 2023). In particular, it can inform ongoing initiatives by industry organizations (e.g. the Frontier Model Forum), standard-setting bodies (e.g. NIST and ISO/IEC), and government authorities (e.g. the AI Office in the EU or DSIT in the UK).

*Questions for further research*. At the same time, the article has left many questions unanswered. In particular, it has not engaged with the more practical question of how exactly frontier AI developers should set up an internal audit function and what that function would do on a day-to-day basis. Future research could adapt the IIA's Global Internal Audit Standards to the context of frontier AI development (IIA, 2024). Once the first developer has set up an internal audit function, scholars could turn towards more empirical research on its effectiveness, drawing from similar research from other industries (e.g. Lenz & Hahn, 2015; Oussii & Taktak, 2018; Carcello et al., 2020). To address some of the challenges involved in transferring best practices from internal IT audits to AI (Table 2), it will be particularly useful to gather empirical evidence. An industry case study similar to the one that Mökander and Floridi (2022) have conducted on ethics-based auditing could be a first step. The article has also not made any concrete policy recommendations. Scholars could review the regulatory concept in other domains—where rules on internal audit are often set out in guidelines, not legislation (e.g. European Banking Authority, 2021)—to learn lessons for frontier AI regulations (Anderljung et al., 2023a; Schuett et al., 2024). Such work could directly inform the regulatory debate on frontier AI regulation in the UK (DSIT, 2023c). Similarly, scholars could suggest specifications on internal audit for the NIST AI Risk Management Framework (NIST, 2023, 2024; Barrett et al., 2023) or harmonized standards for the EU AI Act (Soler Garrido et al., 2023; McFadden et al., 2021; Schuett, 2023b). In general, I wish to encourage more scholars, especially from the field of risk analysis, to contribute to the development of best practices in frontier AI governance (Schuett et al., 2023). Collaborations with researchers from industry may be particularly fruitful, though scholars should pay special attention to concerns around industry capture (Whittaker, 2021).

*An opportunity for frontier AI developers.* Frontier AI developers that have an internal audit function might be perceived as being more responsible. As they transition from startups to more mature companies, many stakeholders (business partners, regulators, etc.) expect them to follow best practices in corporate governance. By setting up an internal audit function, developers would signal that they take risk governance seriously, which might increase their perceived legitimacy.[26] If they do it before regulators mandate action, they would also signal proactivity. Since none of the frontier AI developers seems to have an internal audit function, the first developer would likely get most of the PR benefits. They have an opportunity to be ahead of the curve.

In light of rapid progress in AI research and development, frontier AI developers need to strengthen their risk governance. Instead of reinventing the wheel, they should follow existing best practices. While this might not be sufficient, they should not skip this obvious first step.

## List of abbreviations

| | |
|---|---|
| AGI | Artificial general intelligence |
| AI | Artificial intelligence |
| API | Application programming interface |
| CAE | Chief Audit Executive |
| CEO | Chief Executive Officer |
| CIA | Certified Internal Auditor |
| CRO | Chief Risk Officer |
| DNN | Deep neural network |
| DSIT | UK Department for Science, Innovation and Technology |
| ERM | Enterprise risk management |
| FLOP | Floating point operations |
| FMF | Frontier Model Forum |
| FTE | Full-time employee |
| GPAI | General-purpose AI |
| IIA | Institute of Internal Auditors |
| KYC | Know-your-customer |
| LLM | Large language model |
| LMM | Large multimodal model |
| LTBT | Long-Term Benefit Trust |
| NIST | National Institute of Standards and Technology |
| RL | Reinforcement learning |
| RLHF | Reinforcement learning from human feedback |

---

[26] In the context of external audit, there is a concern that it is more important to an organization's legitimacy that it is *seen* to be audited than that there is any real substance to the audit (Power, 1984). By analogy, the same concern might apply to internal audit.

| | |
|---|---|
| RLAIF | Reinforcement learning from AI feedback |
| RSC | Google DeepMind's Responsibility and Safety Council |
| RSP | Anthropic's Responsible Scaling Policy |
| SAG | OpenAI's Safety Advisory Group |
| SLSA | Supply Chain Levels for Software Artifacts |
| SSDF | NIST Secure Software Development Framework |

*Forthcoming in Risk Analysis* 35

*Forthcoming in Risk Analysis* 41*support of the AI Act*. European Commission, Joint Research Centre. https://doi.org/10.2760/5847

Srivastava, A., Rastogi, A., Rao, A., Shoeb, A. A. M., Abid, A., Fisch, A., Brown, A. R., Santoro, A., Gupta, A., Garriga-Alonso, A., Kluska, A., Lewkowycz, A., Agarwal, A., Power, A., Ray, A., Warstadt, A., W. Kocurek, A., Safaya, A., Tazarv, A., … Wu, Z. (2022). *Beyond the Imitation Game: Quantifying and extrapolating the capabilities of language models*. arXiv. http://arxiv.org/abs/2206.04615

Stafford, T., Deitz, G., & Li, Y. (2018). The role of internal audit and user training in information security policy compliance. *Managerial Auditing Journal, 33*(4), 410–424. https://doi.org/10.1108/MAJ-07-2017-1596

Steinbart, P. J., Raschke, R. L., Gal, G., & Dilla, W. N. (2012). The relationship between internal audit and information security: An exploratory investigation. *International Journal of Accounting Information Systems, 13*(3), 228–243. https://doi.org/10.1016/j.accinf.2012.06.007

Stewart, J., & Subramaniam, N. (2010). Internal audit independence and objectivity: Emerging research opportunities. *Managerial Auditing Journal*, *25*(4), 328–360. https://doi.org/10.1108/02686901011034162

Stoel, D., Havelka, D., & Merhout, J. W. (2012). An analysis of attributes that impact information technology audit quality: A study of IT and financial audit practitioners. *International Journal of Accounting Information Systems, 13*(1), 60–79. https://doi.org/10.1016/j.accinf.2011.11.001

Taleb, N. N. (2007). *The black swan: The impact of the highly improbable*. Random House.

Tetlock, P. E. (2005). *Expert political judgment: How good is it? How can we know?* Princeton University Press.

Tetlock, P. E., & Gardner, D. (2015). *Superforecasting: The art and science of prediction*. Penguin Random House.

Thekdi, S. A., & Aven, T. (2022). Risk analysis under attack: How risk science can address the legal, social, and reputational liabilities faced by risk analysts. *Risk Analysis, 43*(6), 1212–1221. https://doi.org/10.1111/risa.13984

Trager, R., Harack, B., Reuel, A., Carnegie, A., Heim, L., Ho, L., Kreps, S., Lall, R., Larter, O., Ó hÉigeartaigh, S., Staffell, S., & Villalobos, J. J. (2023). *International governance of civilian AI: A jurisdictional certification approach*. arXiv. https://arxiv.org/abs/2308.15514

Turner, A. M., Smith, L., Shah, R., Critch, A., & Tadepalli, P. (2023). *Optimal policies tend to seek power*. arXiv. https://arxiv.org/abs/1912.01683

Turner, A. M., & Tadepalli, P. (2022). *Parametrically retargetable decision-makers tend to seek power*. arXiv. https://arxiv.org/abs/2206.13477

Urbina, F., Lentzos, F., Invernizzi, C., & Ekins, S. (2022). Dual use of artificial-intelligence-powered drug discovery. *Nature Machine Intelligence*, *4*(3), 189–191. https://doi.org/10.1038/s42256-022-00465-9

Villalobos, P., Sevilla, J., Heim, L., Besiroglu, T., Hobbhahn, M., & Ho, A. (2024). *Will we run out of data? An analysis of the limits of scaling datasets in machine learning*. arXiv. https://arxiv.org/abs/2211.04325

Vipra, J., & Korinek, A. (2023). *Market concentration implications of foundation models*. arXiv. https://arxiv.org/abs/2311.01550

Wang, P. (2019). On defining artificial intelligence. *Journal of Artificial General Intelligence*, *10*(2), 1–37. https://doi.org/10.2478/jagi-2019-0002